\documentclass[conference]{IEEEtran}
\usepackage{cite}
\usepackage{amsmath,amssymb,amsfonts}
\usepackage{breqn}
\usepackage{mathtools}
\usepackage{algorithmic}
\usepackage{textcomp}
\usepackage{xcolor}

\usepackage{booktabs} 
\usepackage{etoc} 
\usepackage{multirow}
\usepackage{makecell}
\usepackage{enumitem}

\usepackage{graphicx} 
\usepackage[export]{adjustbox}
\usepackage{subfig}
\usepackage{epstopdf}
\DeclareGraphicsExtensions{.pdf,.png,.eps,.jpg,.tif,.tiff,.ps}
\graphicspath{{img/temp_images//}{img//}}

\usepackage[hyphens]{url}
\usepackage{hyperref} 
\hypersetup{breaklinks=true} 
\usepackage[nameinlink,capitalize]{cleveref}

\usepackage{xspace}
\usepackage[colorinlistoftodos,textsize=tiny]{todonotes}
\definecolor{navy}{rgb}{0.1, 0.1, 0.8}
\definecolor[named]{gray}{rgb}{0.4, 0.4, 0.4}
\definecolor[named]{olive}{rgb}{0.1, 0.5, 0.1}
\definecolor[named]{ruby}{rgb}{0.8, 0.1, 0.3}
\definecolor{darkpastelgreen}{rgb}{0.01, 0.75, 0.24}
\definecolor{celestialblue}{rgb}{0.29, 0.59, 0.82}
\definecolor{coral}{rgb}{1.0, 0.5, 0.31}
\definecolor{Goldenrod}{rgb}{0.8,0.8,0}

\newcommand{\eat}[1]{}

\newcommand{\revA}[1]{{#1}}
\newcommand{\verify}[1]{{}}

\newcommand{\NOTE}[2]{ }
\newcommand{\TODO}[2]{\xspace}
\newcommand{\nb}[1]{}
\newcommand{\note}[1]{\xspace}
\newcommand{\mar}[1]{\xspace}

\newcommand{\ND}[1]{\xspace}

\def\BibTeX{{\rm B\kern-.05em{\sc i\kern-.025em b}\kern-.08em
    T\kern-.1667em\lower.7ex\hbox{E}\kern-.125emX}}
\begin{document}
\etocdepthtag.toc{mtchapter}

\newcommand{\titlename}{Predicting Skill Shortages in Labor Markets:\\A Machine Learning Approach}
\title{\titlename}

\makeatletter
\newcommand{\linebreakand}{%
  \end{@IEEEauthorhalign}
  \hfill\mbox{}\par
  \mbox{}\hfill\begin{@IEEEauthorhalign}
}
\makeatother

\author{\IEEEauthorblockN{
Nik Dawson\IEEEauthorrefmark{1}\IEEEauthorrefmark{2}\IEEEauthorrefmark{3},
Marian-Andrei Rizoiu\IEEEauthorrefmark{4}\IEEEauthorrefmark{5},
Benjamin Johnston\IEEEauthorrefmark{2},
Mary-Anne Williams\IEEEauthorrefmark{2}}
\IEEEauthorblockA{
\IEEEauthorrefmark{2}\textit{Centre of Artificial Intelligence, University of Technology Sydney},\\ 
\IEEEauthorrefmark{3}\textit{OECD Future of Work Research Fellow},\\
\IEEEauthorrefmark{4}\textit{Faculty of Engineering \& IT, University of Technology Sydney}, \IEEEauthorrefmark{5}\textit{CSIRO's Data61},
Sydney, Australia\\
Email: nikolas.j.dawson@student.uts.edu.au
\IEEEauthorrefmark{1}Corresponding author.
}}

\maketitle
\thispagestyle{plain}
\pagestyle{plain}

\begin{abstract}
Skill shortages are a drain on society.
They hamper economic opportunities for individuals, slow growth for firms, and impede labor productivity in aggregate. 
Therefore, the ability to understand and predict skill shortages in advance is critical for policy-makers and educators to help alleviate their adverse effects. 
This research implements a high-performing Machine Learning approach to predict occupational skill shortages.
In addition, we demonstrate methods to analyze the underlying skill demands of occupations in shortage and the most important features for predicting skill shortages. 
For this work, we compile a unique dataset of both Labor Demand and Labor Supply occupational data in Australia from 2012 to 2018.
This includes data from 7.7 million job advertisements (ads) and 20 official labor force measures. 
We use these data as explanatory variables and leverage the XGBoost classifier to predict yearly skills shortage classifications for 132 standardized occupations. 
The models we construct achieve macro-F1 average performance scores of up to 83 per cent.
Our results show that job ads data and employment statistics were the highest performing feature sets for predicting year-to-year skills shortage changes for occupations.
We also find that features such as `Hours Worked', years of `Education', years of `Experience', and median `Salary' are highly important features for predicting occupational skill shortages.
This research provides a robust data-driven approach for predicting and analyzing skill shortages, which can assist policy-makers, educators, and businesses to prepare for the future of work.
\end{abstract}

\begin{IEEEkeywords}
Big Data, Data Science, Skill Shortages, Job Advertisements, Labor Economics
\end{IEEEkeywords}

\section{Introduction}
\label{sec:intro}
In January 2019, Andrew Penn, the CEO of Telstra -- Australia's largest Telecommunications company -- announced that the company will be expanding its new `Innovation and Capability Center' in Bangalore, India. This will create approximately 300 Network and Software Engineering jobs, with the potential for more~\cite{Pearce2019}.
Penn cited `skill shortages' as the main reason for this outsourcing decision:

\begin{quote}
	``We need these capabilities now, but the fact is we cannot find in Australia enough of the skills that we need on the scale that we need them, such as software engineers. Why? There simply are not enough of them. The pipeline is too small.'' \cite{CEDA}
\end{quote}

This coincides with Telstra announcing a goal net reduction of 8,000 jobs by 2022 (mainly in Australia), as the company seeks to automate labor tasks and simplify processes~\cite{Chalmers2019-ha}.
While an isolated example, the evolving labor demands of Telstra highlights both the opportunity costs of skill shortages and the precariousness of workers' security to automation and globalization. 
As a result of these claimed skill shortages, the Australian labor market will not enjoy the benefits afforded by 300 highly skilled jobs -- benefits that materialize in greater economic activity, labor productivity, and economic competitiveness.
This is not specific to just Telstra or Australia, skill shortages burden most labor markets to varying extents~\cite{brunello2019skill}. Their impacts limit employment opportunities for individuals, impede technology adoption and investment by firms, and hamper labor productivity in aggregate.

In this work, we focus on three open problems relating to skill shortages at the occupational level. The first problem relates to analyzing the underlying skills of occupations known or suspected to be in shortage. Skills enable workers to complete labor tasks that are required by jobs. Therefore, analyzing the demand and relative importance of skills within occupations provides granular insights into which skills should be developed and prioritized for occupations in shortage. This can help to inform policy-makers, educators, and individual job-seekers. However, most approaches to determining skill importance within an occupation have relied on \textit{ad hoc} aggregations of job advertisements (ads) or résumé data~\cite{lee2005analysis, Gardiner2018-dt}. While simple frequency counts can provide useful proxies for demand, such methods do not normalize for highly common skills and can therefore yield distorted views of skill importance within occupations. So, the question is \textbf{(1) can we determine which skills are most important for occupations in shortage while accounting for highly common skills?}

The second open problem is concerned with predicting occupational skill shortages. While the adverse effects of skill shortages have been well-documented~\cite{brunello2019skill,haskel1993skill,OECD2019Skills}, predicting skill shortages is difficult. 
Even more challenging is predicting temporal changes to the skill shortage status of an occupation. 
For example, accurately predicting whether an occupation will shift from being classified as \textit{Not in Shortage} in one time period to \textit{In Shortage} the next. 
These difficulties reflect the lack of consensus around which variables are most predictive of skill shortages and the limited available data classifying occupational shortages.
The question is, therefore, \textbf{(2) can we leverage modern Data Science and Machine Learning techniques to predict occupational skill shortages?}

The third open problem relates to understanding which variables are most predictive of skill shortages. While many studies have examined the presence of skill shortages in labor markets~\cite{haskel1993skill, junankar2009, cedefop2015shortages}, there remains a lack of understanding about which factors contribute most to occupational shortages. This leads to the final question: by building predictive models, \textbf{(3) can we uncover which variables are most important for predicting skill shortages at the occupational level?}

We address each of the above-stated questions by leveraging both labor demand and labor supply data. With regards to the first research question, we use a rich dataset of 7,697,568 job ads in Australia to analyze the underlying skill demands of `Data Scientists', an occupation shown to be in shortage in Australia~\cite{Dawson2019}, the UK~\cite{Blake2019-ha}, and the US~\cite{Markow2017-hu}. Here, we compare two different methods to assess the top temporal skill demands of `Data Scientists'. We highlight the shortcomings of \textit{ad hoc} skill counts and illustrate an alternative method that captures specialized and emerging skills within an occupation.

We address the second research question by constructing a supervised Machine Learning model framework to predict skills shortage classifications at the occupational level one year into the future.
These binary classification models are built using eXtreme Gradient Boosting (XGBoost), a scalable Machine Learning system for tree boosting~\cite{Chen2016}.
We incorporate labor demand and labor supply occupational data from Australia as input, which are organized and matched according to the official Australian occupational standards.
On the labor demand side, we again use job ads data from the aforementioned dataset, spanning from 2012-01-01 to 2018-12-31.
For the labor supply side, we use `Detailed Labor Force' data from the Australian Bureau of Statistics over the same time period~\cite{Australian_Bureau_of_Statistics2019-sv}.
Lastly, the `ground-truth' (or predictive variable) is taken from the longitudinal list of occupational shortages, recorded by the Australian Federal Department of Education, Skills and Employment~\cite{Department_of_Education_Skills_and_Employment_Australian_Government2019-qo}. These official skill shortage classifications directly inform national and state policies in the areas of education, training, employment and skilled immigration. Further detail on the data is discussed in \cref{sec:data-methods}.

Lastly, we address the third research question by extracting the feature importance data generated from the above prediction model. This sheds light on which variables are most important for predicting the skill shortage status of an occupation. Importantly, we find empirical evidence that `Hours Worked', `Education', `Years of Experience', and `Salary' are the most important features for predicting occupational skills shortages.
This supports evidence from Labor Economics where workers in occupations experiencing skill shortages tend to have higher work intensity and longer work hours~\cite{healy2015, Richardson2007-vl}. Similarly, employers attempt to overcome skill shortages and meet labor demands by lowering education requirements, experience demands, and increase salary levels to attract a greater pool of candidates~\cite{healy2015, brunello2019skill, Dawson2019}. These variables prove to be predictive features (see \cref{subsec:feature-importance}).

\textbf{The main contributions of this work are the following:}
\begin{itemize}
    \item We compare two methods to \textbf{analyze the underlying skill demands of occupations in shortage and detect emerging skills}, using `Data Scientists' as the example;
    \item We implement a \textbf{data-driven modeling framework to predict temporal skill shortages of occupations};
    \item Lastly, we \textbf{analyze the feature importance data from the prediction models to identify which variables are most predictive of skill shortages}.
\end{itemize}

\section{Related Work \& Limitations} \label{related-work}

We structure this discussion of the related work into two areas.
First, in \cref{subsec:measure-labor-shortage}, we visit work dealing with measuring skill shortages.
Second, in \cref{subsec:economic-costs}, we investigate the economic costs of skill shortages.

\subsection{Measuring Labor Shortages}
\label{subsec:measure-labor-shortage}

\textbf{The broader problem.}
Skill shortages occur when the labor demand for specific skills exceed the supply of workers who possess those skills at a prevailing market wage~\cite{junankar2009, healy2015}. 
Skill shortages can be considered a subset of the broader problem of `Skills Mismatch'. 
At the macro-level, skills mismatch refers to the disequilibrium of aggregate supply and demand of labor skills, usually with reference to a specific geographic unit~\cite{brunello2019skill}. 
Skill shortages are one scenario of skills mismatch and occur when the demand for specific skills exceed the available supply of workers at real wage rates. Conversely, `Skill Surpluses' are caused by an excess of skill supply~\cite{quintini2011right}.
That is, there are more workers who possess specific skills that the labor market demands on aggregate. 
Therefore, skill shortages are usually calculated as a component of measuring skill mismatches.

For a discussion on the factors that cause skill shortages, please refer to the \revA{online appendix}~\cite{appendix}.

\textbf{Measures using surveys.}
Skill shortages are typically measured at the firm-level through the use of surveys to examine the extent of unfilled and hard-to-fill vacancies~\cite{mcguinness2017}.
A shortcoming of this approach is that skill shortages can be overstated and such surveys are often unrepresentative. 
For instance, employers may claim an occupation to be \textit{In Shortage} but the underlying cause could be their own inability to offer a sufficient wage-level, attractive working conditions, or a desirable location. 
These micro-level factors can distort the presence of genuine skill shortages, where employers extrapolate their firm-specific challenges as macro-level issues~\cite{cedefop2015shortages, appendix}.

\textbf{Use of indirect measures.} \label{subsec:indirect-measures}
To differentiate between perceived and genuine skill shortages, other studies have complemented survey results with indirect measures, such as wage growth, employment growth, vacancy rates, and work intensity. 
The rationale underlying these approaches is that occupations experiencing skill shortages are typically characterised by wage premiums, greater employment growth, growing vacancy rates, and higher work hours and levels of overtime~\cite{brunello2019skill}.
The OECD implemented such indirect measures in concert with employer surveys to construct a series of indicators and composite indexes on skills for employment, including skill shortages~\cite{OECDSkills2017}.
The `World Indicators of Skills for Employment' (WISE) database calculates an occupational indicator of skill shortages based on wage growth, employment growth, and growth in the hours worked~\cite{Oecd2015WISE}.
Next, this indicator is transformed into a composite skill index that uses the O*NET database~\cite{ONET} to map occupations into groups of skills and tasks. 
This allows for international comparability between OECD countries for skills challenges and performances, including the extent of skill shortages.

Other approaches have used indicators from job ads data to assess skill shortages. 
Dawson et al.~\cite{Dawson2019} analyzed a large temporal dataset of online job ads to detect skill shortages of Data Science and Analytics occupations in Australia. The authors use a range of indicators to evaluate the presence and extent of skill shortages, such as posting frequency, salary levels, educational requirements, and experience demands. 
They contend that 
occupations experiencing high posting growth \revA{appear} volatile and \revA{their posting frequencies are} difficult to predict.
Given that high and growing posting frequency is often used as a proxy for high labor demand for occupations, the authors argue that high error metrics, combined with the other indicators, can help detect skill shortages. 
\revA{In this work,} we use \revA{the} labor demand features features \revA{proposed} in Dawson et al.~\cite{Dawson2019} \revA{to build a skill shortage classifier}.
\revA{For completeness reasons, we describe these features in the online appendix~\cite{appendix}}.

\textbf{The current work.}
The present work takes a data-driven machine learning approach to measure and predict skill shortages.
We leverage a set of recently proposed labor demand features extracted from job ads data~\cite{Dawson2019}, together with official labor supply features to build a machine learning model that classifies whether an occupation is in shortage. In addition, we analyze the relative importance of these features.

\subsection{Economic Costs of Skill Shortages}
\label{subsec:economic-costs}
The costs of skill shortages can be significant and manifest at both micro and macro-levels of economies. They affect individuals, firms, and aggregate markets. 

\textbf{Individual-level.}
Skill shortages can negatively affect earnings and reduce development opportunities for workers. 
Markets experiencing skill shortages can force individuals to accept less desirable and insecure work. 
In 2011, Quintini~\cite{quintini2011right} analyzed household survey data from the European Community Household Panel to investigate the effects of qualification mismatch on earning. 
Quintini found that `over-qualified' individuals earn approximately 3\% less than individuals with the same occupations but who have been appropriately matched. 
The presence of skill shortages exacerbates the inefficient allocation of labor, which can negatively affect the earnings and employment opportunities for individuals.

\textbf{Firm-level.}
Several studies have examined the implications of skill shortages on firm-level productivity and all concluded that skill shortages negatively impact firm-level productivity~\cite{bennett2009, forth2006ict, tang2005, haskel1996}.
In a study using the Australian Business Longitudinal Database, Healy et al.~\cite{healy2015} found that most Australian firms respond to skill shortages through longer working hours and higher wages for occupations experiencing in shortage. 
Significantly, we found that the `Hours Worked' and `Salary' levels were among the most important features for predicting skill shortages, seen in \cref{subsec:feature-importance}.
However, there is evidence to suggest that such skill shortages are usually short-lived. Further research analyzed the existence of skill shortages in German firms and concluded that while their effects can be acute, they are typically a temporary and short-term phenomena~\cite{bellmann2014}.

\textbf{Macroeconomic-level.}
Lastly, the economic costs of skill shortages accumulate to macroeconomic effects. Frogner~\cite{frogner2002skills} uses data from the Employers Skill Survey to identify the negative impacts of skill shortages on productivity, Gross Domestic Product, employment levels, and wage earnings. 
From the perspective of private investment, Nickell et al.~\cite{nickell1997} calculates that a 10\% increase in firms reporting skill shortages decreases private investment by 10\% and Research \& Development investment by 4\%. 
The inefficient allocation of resources caused by skill shortages therefore hampers productivity, which can compromise macroeconomic growth.

\textbf{The current work} proposes a method to predict in advance skill shortages and better understand their contributing factors.
These methods and results could in turn be used by policy-makers, educators, and companies to prepare for and alleviate the negative impacts of skill shortages.

\section{Data and Methods}
\label{sec:data-methods}

In this section, we first detail the data sources and the constructed labor demand and labor supply features (\cref{subsec:data-sources-features}).
We then outline two methods to assess skill importance for occupations classified as in shortage (\cref{subsec:skill-importance-methods}.
Last, we detail the prediction model setup and evaluation (\cref{subsec:predictive-model}).
 
\subsection{Data sources and constructed features}
\label{subsec:data-sources-features}

In this work, we employ both labor demand and labor supply data as explanatory variables (features, henceforth) \revA{to predict} occupational skill shortages.
The dataset we construct relates to occupations in Australia during the period 2012-2018. 
\revA{Due to space constraints, the table summarizing all the onstructed features is shown in the}
\revA{online appendix}~\cite{appendix}.

\textbf{Labor demand features.}
For labor demand, we have used job ads data, which was generously provided by Burning Glass Technologies\footnote{BGT is a leading vendor of online job ads data: \texttt{https://www.burning-glass.com/}} (BGT). 
The data has been collected via web scraping and systematically processed into structured formats. 
The dataset consists of detailed information on individual job ads, such as location, salary, employer, educational requirements, experience demands, and more. 
Each job ad is also categorized into its relevant occupational classification. 
We build upon the results of Dawson et al.~\cite{Dawson2019} and we incorporate a range of the engineered job ads indicators that the authors found predictive of labor shortages, as discussed in \cref{subsec:indirect-measures}.

While data from BGT integrates multiple online sources and arguably represents the most comprehensive repository of job ads data, it is argued that \revA{online job ads} are an incomplete representation \revA{of labor demand}~\cite{Carnevale2014-xc}, for two reasons.
\revA{First,} some employers continue to use traditional forms of advertising for vacancies, such as newspaper classifieds, their own hiring platforms, or recruitment agency procurement. 
\revA{Second,} job ads data also over-represent occupations with higher-skill requirements and higher wages, colloquially referred to as `white collar' jobs~\cite{Carnevale2014-xc}.
\revA{These are limitations of the current work, discussed in \cref{sec:conclusion}.}

\textbf{Labor supply features.}
The labor supply data used for this research comes from the `Quarterly Detailed Labor Force' statistics by the Australian Bureau of Statistics (ABS)~\cite{Australian_Bureau_of_Statistics2019-sv}.
This consists of statistics on employment levels, unemployment, underemployment, hours worked and others. 
As the labor supply statistics are measured quarterly, the yearly average for each feature was calculated to match the skills shortage target variable, which is measured in yearly periods \revA{(presented next)}.

\textbf{Skill shortages ground-truth.}
The ground-truth comes from the `Historical List of Skill Shortages in Australia', measured by the Australian Federal Department of Education, Skills and Employment (DESE, henceforth)~\cite{Department_of_Education_Skills_and_Employment_Australian_Government2019-qo}.
For over three decades, the DESE has conducted ongoing skills shortage research in Australia. 
Their research aims to identify shortages for skilled occupations where long lead times for training means that such shortages cannot be addressed immediately. 
The DESE tracks 132 occupations nationally, and they also provide more detailed analyses on select occupations at the State and Territory levels. 
To assess skill shortages, the DESE survey employers every year, called the `Survey of Employers who have Recently Advertised' (SERA). 
The SERA collects both qualitative data from employers and recruitment professionals, and quantifiable data on employers' recruitment experiences~\cite{DoEMethodology}.
The output of this DESE activity is that, for every year, each of the 132 tracked occupations is classified as \textit{In Shortage} or \textit{Not In Shortage} at the national-level. 
The results of these classifications have direct implications for education, training, employment and migration policies.

There are, however, \revA{five} important limitations of the DESE's methodology for measuring skill shortages. 
First, the DESE acknowledge that the survey is not a statistically valid sample of Australia's labor market. 
Second, there are inherent limitations of determining skill shortages from surveying employers, as discussed in \cref{subsec:measure-labor-shortage}. 
Nonetheless, the ABS evaluated the methodology and found that it was "appropriate for its purpose"~\cite{DoEMethodology}. \textbf{To our knowledge, this dataset is the most reliable source of occupational skill shortages that is publicly available in Australia.}
Third, the surveyed occupations in this research are biased towards the occupational classes of `Technicians and Trades' workers and `Professionals'.
Forth, the dataset is imbalanced with a greater number of occupations classified as \textit{Not in Shortage}. 
Fifth and finally, there are inherent limitations that emerge from analyzing jobs using standardized occupational taxonomies. Specifically, official occupational classifications are usually static taxonomies that are rarely updated and slow to adapt to changing labor dynamics. 
This research uses the official Australian and New Zealand Standard Classification of Occupations (ANZSCO)~\cite{ANZSCO2013}.
While other more adaptive taxonomies exist, ANZSCO remains the official taxonomy and is the measurement standard used for all data in this research.

\subsection{\revA{Quantify} skill importance for occupations}
\label{subsec:skill-importance-methods} 
Here, we detail two approaches for determining relative levels of skill importance for an occupation known or suspected to be in shortage. 
We \revA{exemplify both} methods in \cref{subsec:data-scientists-skill-importance} using job ads classified as `Data Scientists' from 2015-2019 in Australia, as this occupation has be shown to be in shortage during this period~\cite{Dawson2019, Deloitte_Access_Economics2018-dh}. 
Analyzing the underlying skills of occupations in shortage is important as it provides granular details on which skills should be targeted to help alleviate occupational shortages. This assists policy-makers, educators, and job-seekers to prioritize the development of specific skills to help meet evolving labor demands.

\textbf{Posting frequency as a proxy for demand.} 
The \revA{proxy most widely used in literature}~\cite{Carnevale2014-xc, Markow2017-hu, Blake2019-ha} for skill importance is skill frequency -- \revA{i.e. count how many times a skill appears in the job ads associated with a given occupation during a predetermined period of time}. 
While skill \revA{frequency} can provide some indication of labor demand (i.e. higher skill counts being indicative of higher demand), it fails to normalize for skills that are demanded by all or most jobs. 
This does not necessarily reveal which skills are more or less important to a given occupation, as some skills generalize across all occupations at high frequencies (for e.g. `Communication Skills' and `Teamwork'). This leads to an alternate method for assessing skill importance within occupations.

\textbf{Normalized skill importance.} 
Here, we use an established measure called `Revealed Comparative Advantage' ($RCA$) that has been applied across a range of disciplines, such as trade economics~\cite{Hidalgo2007-qk,Vollrath1991-kr}, identifying key industries in nations~\cite{Shutters2016-fe}, and detecting the labor polarization of workplace skills~\cite{Alabdulkareem2018-jl}. 
$RCA$ measures the importance of a skill in a job ad, relative to the total share of demand for that skill in all job ads. 
\revA{Formally, the $RCA$  for skill $s$ and the job ad $j$ is}:
\begin{equation*} \label{eq:RCA}
  RCA(j, s) = \frac{x(j, s) / \mathop{\sum}\limits_{s'\in \mathcal{S}}x(j, s')}
  {\mathop{\sum}\limits_{j'\in J}x(j', s) / \mathop{\sum}\limits_{j'\in \mathcal{J},s'\in \mathcal{S}}x(j', s')}
\end{equation*}
\noindent
where $x(j,s) = 1$ when the skill $s$ is required for job $j$, and $x(j,s) = 0$ otherwise;
$\mathcal{S}$ is the set of all distinct skills, and $\mathcal{J}$ is the set of all job ads in our dataset.
$RCA(j, s) \in \left[ 0, \mathop{\sum}\limits_{j'\in J,s'\in S}x(j', s') \right], \forall j, s$, and the higher $RCA(j, s)$ the higher is the comparative advantage (or importance) that $s$ is considered to have for $j$.
Visibly, $RCA(j, s)$ decreases when the skill $s$ is more common (i.e. when $\mathop{\sum}\limits_{j'\in J}x(j', s) $ increases), or when many other skills are required for the job $j$ (i.e. when $\mathop{\sum}\limits_{s'\in S}x(j, s')$ increases).
\revA{Therefore,}
$RCA$ adjusts for the biases that emerge from high-occurring skills across all jobs, while maximizing the skill-level information within individual jobs.

We \revA{compute} skill importance weights at the occupational level $W_{s, o}$ -- i.e. \revA{how important is a particular} skill $s$ in the occupation $o$ for year $t$ -- as the mean $RCA$ for skill $s$ in job ads pertaining to occupation $o$ (denoted as $J_{o}$):
\begin{equation*} \label{eq:SSI}
    W_{s, o} = \frac{1}{|J_{o}|}\sum _{j\in J_{o}, j\in t} RCA(j, s)
\end{equation*}

\revA{As a last step,} we sort the skills by $W_{s, o}$ in descending order, filtering out extremely rare skills \revA{that occur less than five times during a year}. 
This returns a list of top skills that can be interpreted as the most important to occupation $o$ for year $t$, adjusted for high-occurring skills. As is seen in \cref{subsec:data-scientists-skill-importance}, the resulting skills list from this method yields newly emerging and more specific skills than that of posting frequency.

\subsection{\revA{Predictive Setup for} Skill Shortages}
\label{subsec:predictive-model}

\textbf{\revA{Choosing a classification model}.}
In this work, we predict skill shortages by employing XGBoost~\cite{Chen2016} -- an off-the-shelf \revA{classification} algorithm.
XGBoost is an implementation of gradient boosted tree algorithms.
XGBoost has achieved state-of-the-art results on many standard classification benchmarks and is a well established Machine Learning framework~\cite{Orzechowski}.
As an overview, these are Machine Learning techniques that produce prediction models in the form of an ensemble of weak prediction models (here decision trees), by optimizing a differentiable loss function~\cite{chen2014introduction}.
We chose to use XGBoost because it is the currently the state-of-the-art in \revA{both classification and} regression tasks for medium sized amounts of data (i.e. where neural networks cannot be fully deployed).
It also features several advantages that we leverage in our regression task:
it automatically handles missing data values, and supports parallelization of tree construction.

\textbf{Accounting for the temporal inertia of shortage classifications.}
Skill shortages are constantly evolving and labor markets take time to adjust. 
\revA{As a result, skill shortages exhibit strong auto-regressive properties (as can be observed in \cref{sec:results}).
Therefore, we construct models to predict skill shortages which account for these temporal characteristics. }

XGBoost, was not specifically built for time series prediction tasks and \revA{it makes the fundamental assumption that observations are independent}. 
However, XGBoost has been applied for several time series prediction tasks and achieved impressive results~\cite{Zhou2019-gm, Ji2019-jv, Pavlyshenko2016-jf}. 
We also use XGBoost to make predictions on temporal data in this research. 
To account for the temporal nature of skill shortages, we engineer `auto-regressive lagged features' -- i.e. for each feature included in the model, we also include its offset values over a specified number of past periods. 
\revA{In our experiments in \cref{sec:results}, we use two auto-regressive lag periods.}
The inclusion of such auto-regressive lagged features provides each observation with temporal characteristics.

\textbf{\revA{Predicting one year into the future.}} 
The dataset is organised into yearly intervals to match the ground truth.
While \revA{the descriptive features} are available at most three months after the year's end, 
the DESE skills shortage \revA{ground truth is}
often published 12-18 months (or longer) after the reported period.
\revA{Therefore, our models are setup to predict skill shortages one year in advance of the official government release.}

\textbf{Training model hyper-parameters.}
Like most machine learning algorithms, XGBoost has a set of hyper-parameters -- parameters related to the internal design of the algorithm that cannot be fit from the training data. 
The hyper-parameters are usually tuned through search and cross-validation.
In this work, we employ a Randomized-Search~\cite{bergstra2012random} which randomly selects a (small) number of hyper-parameter configurations and performs evaluation on the training set via cross-validation.
We tune the hyper-parameters for each learning fold using 2500 random combinations, evaluated using a 5 cross-validation. 
We also implemented `oversampling' to accommodate for the imbalance between \revA{the \textit{In Shortage} and \textit{Not in Shortage } classes in the ground-truth  (see \cref{subsec:exploratory-data-analysis})}.
This technique involves randomly duplicating observations from the minority class (\textit{In Shortage}) and adding them to the training dataset (see \cite{appendix} for more details).


\textbf{Performance measures.}
Here, we measure the performance of our prediction using three standard Machine Learning performance measures:
precision, recall, and F1. For more details on these metrics, please refer to the \revA{online appendix}~\cite{appendix}.



In our results in \cref{sec:results}, we report the macro-precision, macro-recall and macro-F1, which are the means of the indicators over the two classes.
\revA{This makes sure} that the minority class (here the \emph{In Shortage} class) are not under-represented in the results.

\textbf{Train-test split.}
Consistent with established Machine Learning practices, we separated the dataset into `training' and `testing' sets. 
This split was implemented temporally, with observations from 2012-2016 included in the training dataset, and observations from 2017-2018 included in the testing dataset. 
The training dataset consisted of 660 observations (71\% of total observations) and the testing dataset consisted of 264 observations (29\% of total observations). 
Segmenting the dataset into temporal training and testing sets is done to ensure objectivity in the evaluation process and reflect the temporal nature of the ground-truth.

\begin{figure*}
	\centering
	\setkeys{Gin}{height = 0.28\textheight} 
	\newcommand\myheightA{0.205} 
	\newcommand\myheightB{0.185} 
	\newcommand\mywidthA{0.32}

	\subfloat[]{
		\includegraphics{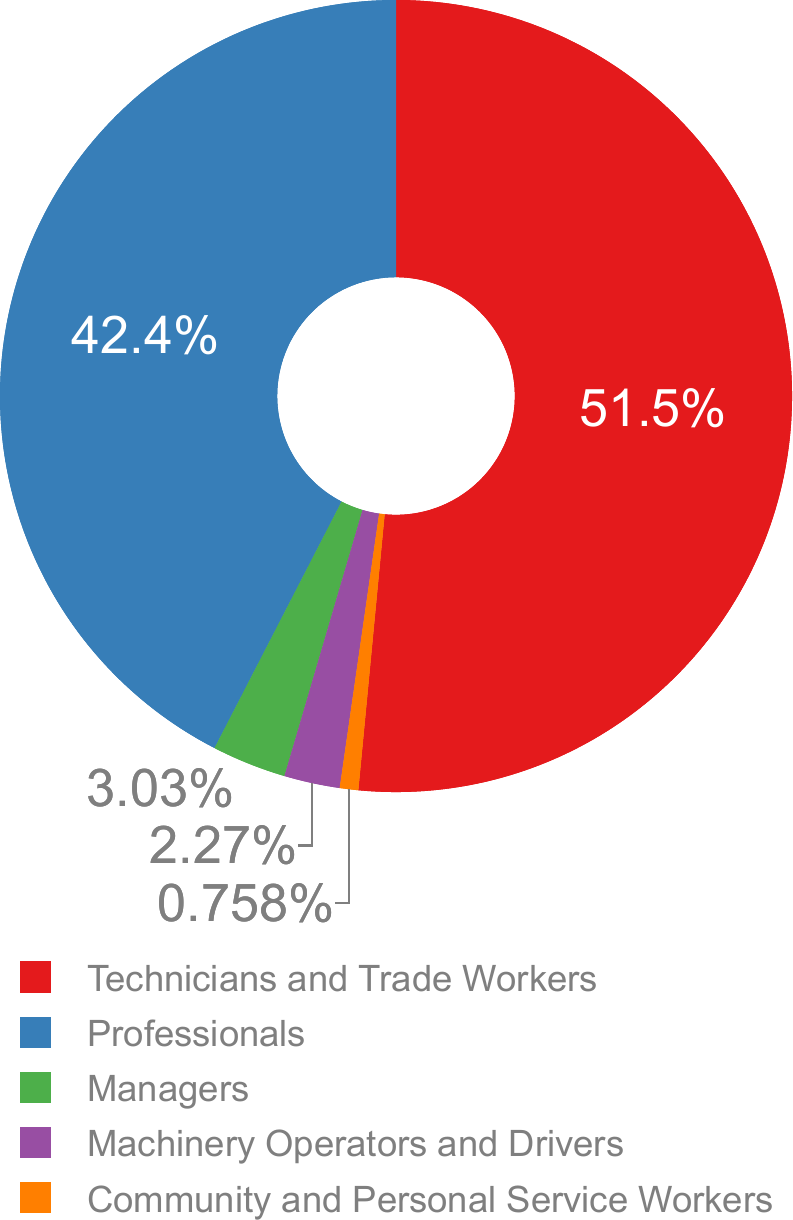}
		\label{subfig:major-group-dist}
	}
	\hfill
	\subfloat[]{
		\includegraphics{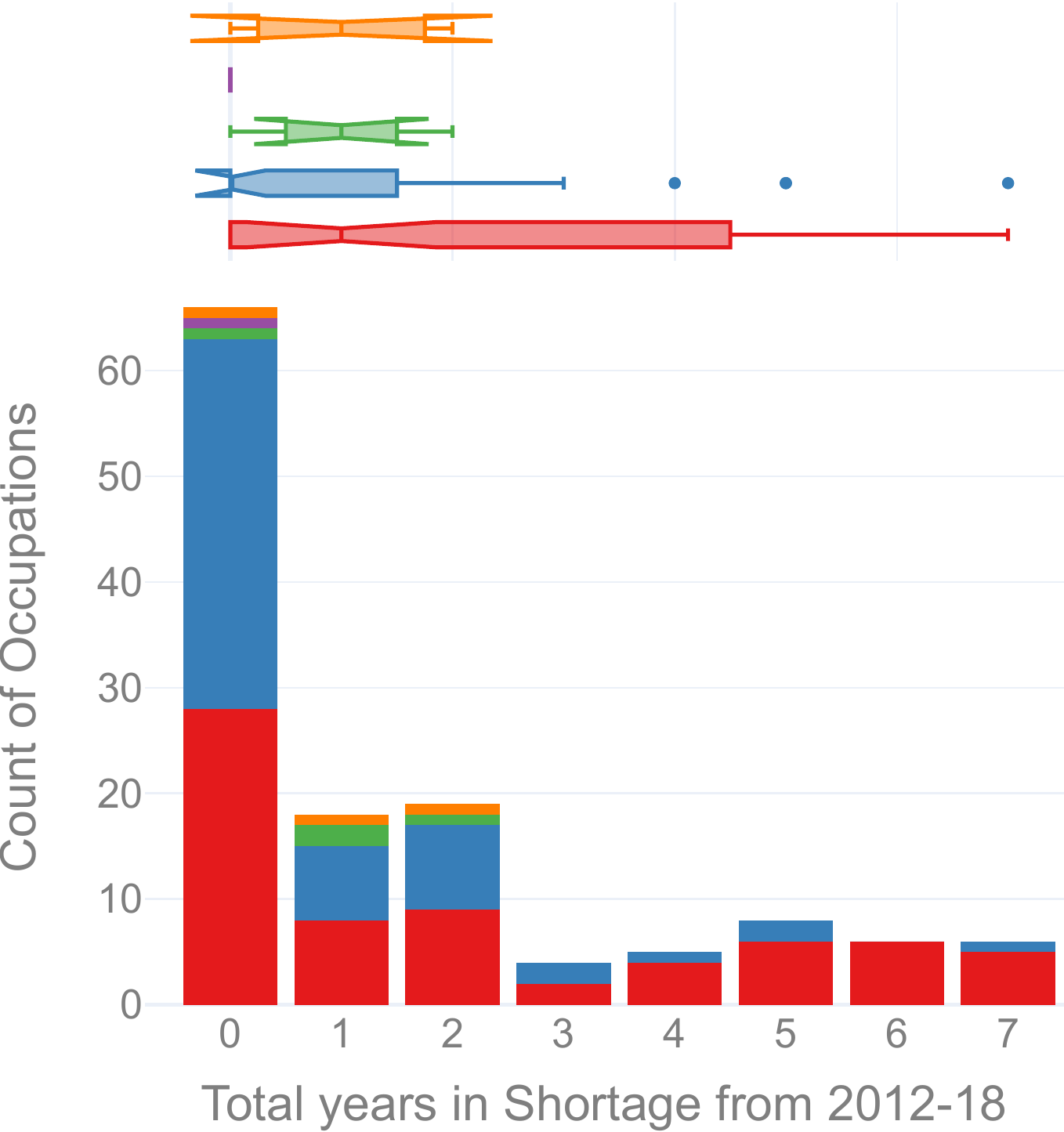}
		\label{subfig:yearly-shortage-dist}
	}
	\hfill
	\subfloat[]{
		\includegraphics{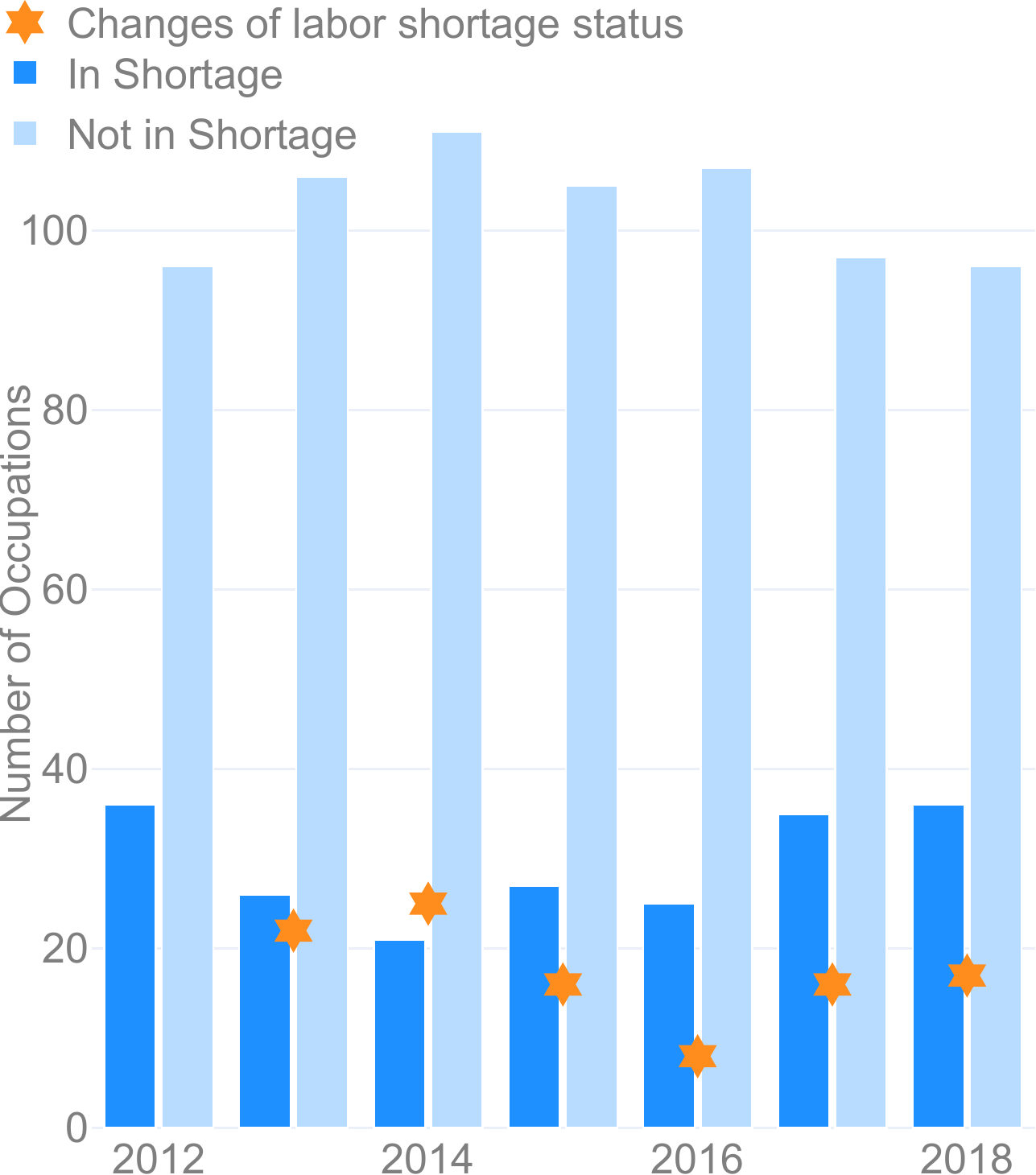}
		\label{subfig:total-shortage-dist}
	}
	\caption{
		\textbf{Overview of Skills Shortage Dataset}:
	    \textbf{(a)} Proportion of occupations represented in dataset by ANZSCO Major Group classes;
	    \textbf{(b)} count of occupations grouped by years \textit{In Shortage};
	    \textbf{(c)} total distribution of occupations classified as \textit{Not in Shortage} (718 total) or \textit{In Shortage} (206 total).
	}
	\label{fig:eda}
\end{figure*}

\begin{figure}[htp]
    \centering
    \includegraphics[width=0.49\textwidth]{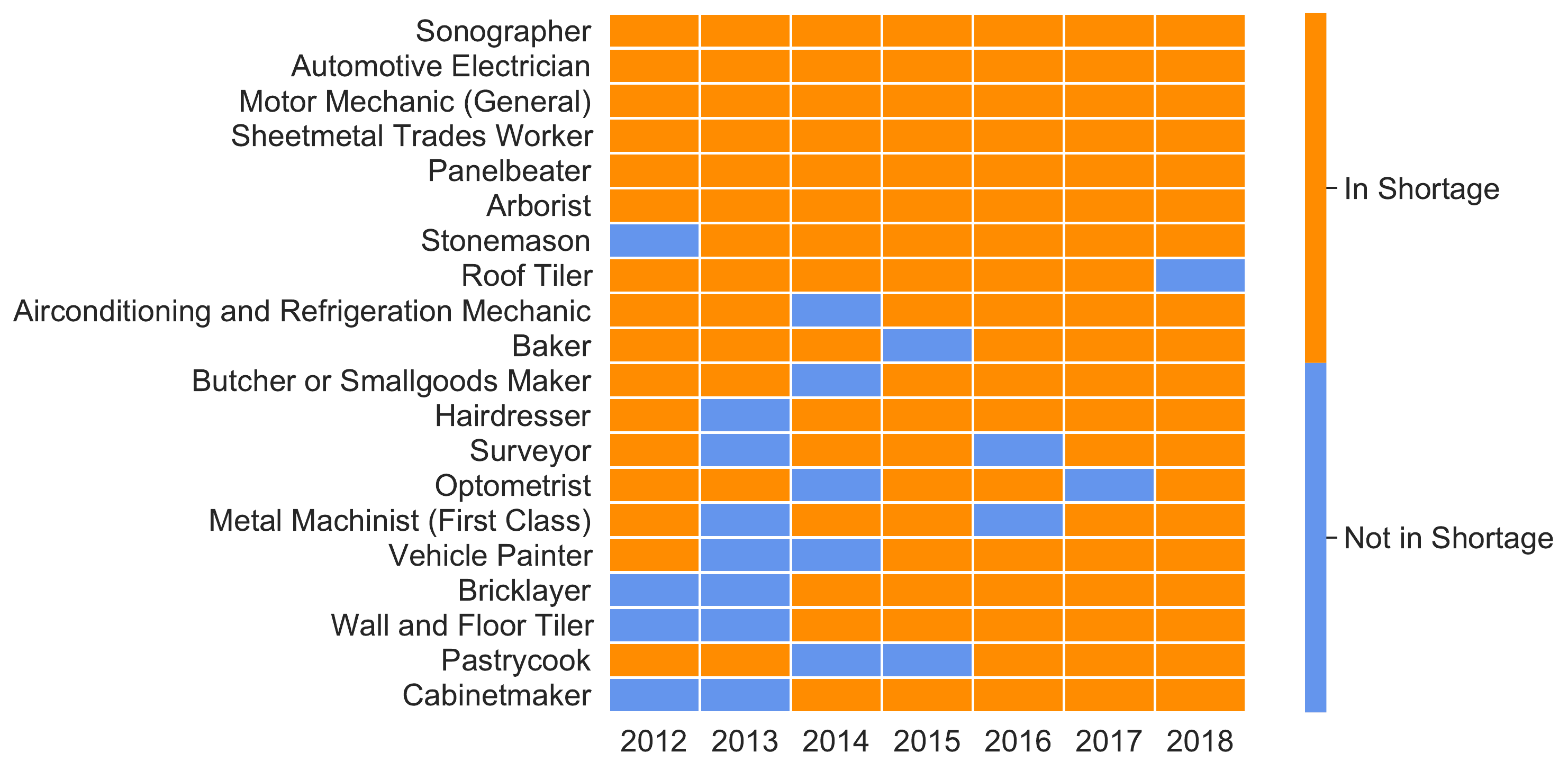}
	\caption{
		Top occupations most \textit{In Shortage} at the ANZSCO 6-digit occupational level.}
	\label{subfig:top-shortage-occs}
\end{figure}

\section{Results}
\label{sec:results}
In this section, we \revA{first} perform an exploratory data analysis of the constructed dataset (\cref{subsec:exploratory-data-analysis}) before showing three sets of results that directly answer our research questions from \cref{sec:intro}.
In the first set of results, we compare two methods to analyze the underlying skill demands of `Data Scientists'. 
Next, we implement Machine Learning models to predict skill shortages of occupations, as outlined in \cref{subsec:predictive-model}. 
Last, we extract and analyze the feature importance data from these models to identify which variables are most predictive of skill shortages.
We incorporate three data sources to construct the dataset that we use for modeling; these data sources include (1) job ads data from BGT, (2) employment statistics from ABS, and (3) occupational skills shortage classifications from the DESE, which are described in \cref{subsec:data-sources-features}.

\subsection{\revA{Profiling the Skills Shortage Prediction dataset}}
\label{subsec:exploratory-data-analysis}
\revA{Here}, we perform an exploratory data analysis and profiling of the dataset.
The purpose is to understand the biases and imbalances introduced during the dataset's construction.

\textbf{\revA{Construct the Skills Shortage Prediction dataset.}}
The compiled dataset describes 132 unique occupations during the period 2012-2018. 
Each row consists of a tuple (occupation, year), and it describes the given occupation during that particular year using its ANZSCO identifiers, the values for each of the descriptive features (described in \cref{subsec:data-sources-features} and the \revA{online appendix}~\cite{appendix}), and the auto-regressive lagged features (described in \cref{subsec:predictive-model}).
\revA{Our resulted dataset contains 924 occupation-year tuples (rows) described by a total of 32 features (excluding lagged feature periods).}
The binary target variable is its shortage status during that year: \textit{In Shortage} or \textit{Not in Shortage}.
In constructing this dataset, we analyzed the auto-correlations within the constructed features, which are presented in the \revA{online appendix}~\cite{appendix}. Unsurprisingly, we found that features from the same or similar categories were strongly correlated, whereas features from different datasets (job ads data and employment statistics) tended to be uncorrelated; the analysis did not yield consequential results.
We next profile the contributed dataset, and we uncover a series of specifics that should be considered during the modeling process.
The Skills Shortage Prediction dataset and code will be made available upon acceptance of the paper.

\textbf{Prevalence of Technicians and Professionals.}
\cref{subfig:major-group-dist} shows that the occupational classes measured by the DESE disproportionately represent `Technicians and Trades' and `Professionals'. 
Collectively, these two major occupational groups account for 94\% of occupations included in the dataset. 
This is higher than the number of workers actually employed in these occupational classes. 
For instance, the ABS indicates that `Professionals' represent approximately 24\% of employment in Australia~\cite{Australian_Bureau_of_Statistics2019-sv}. 
\revA{The bias and validity of the ground truth are discussed in \cref{subsec:data-sources-features}.}

\begin{figure*}
	\centering
	\newcommand\myheightA{0.315} 

	\subfloat[]{
		\includegraphics[height = \myheightA\textheight]{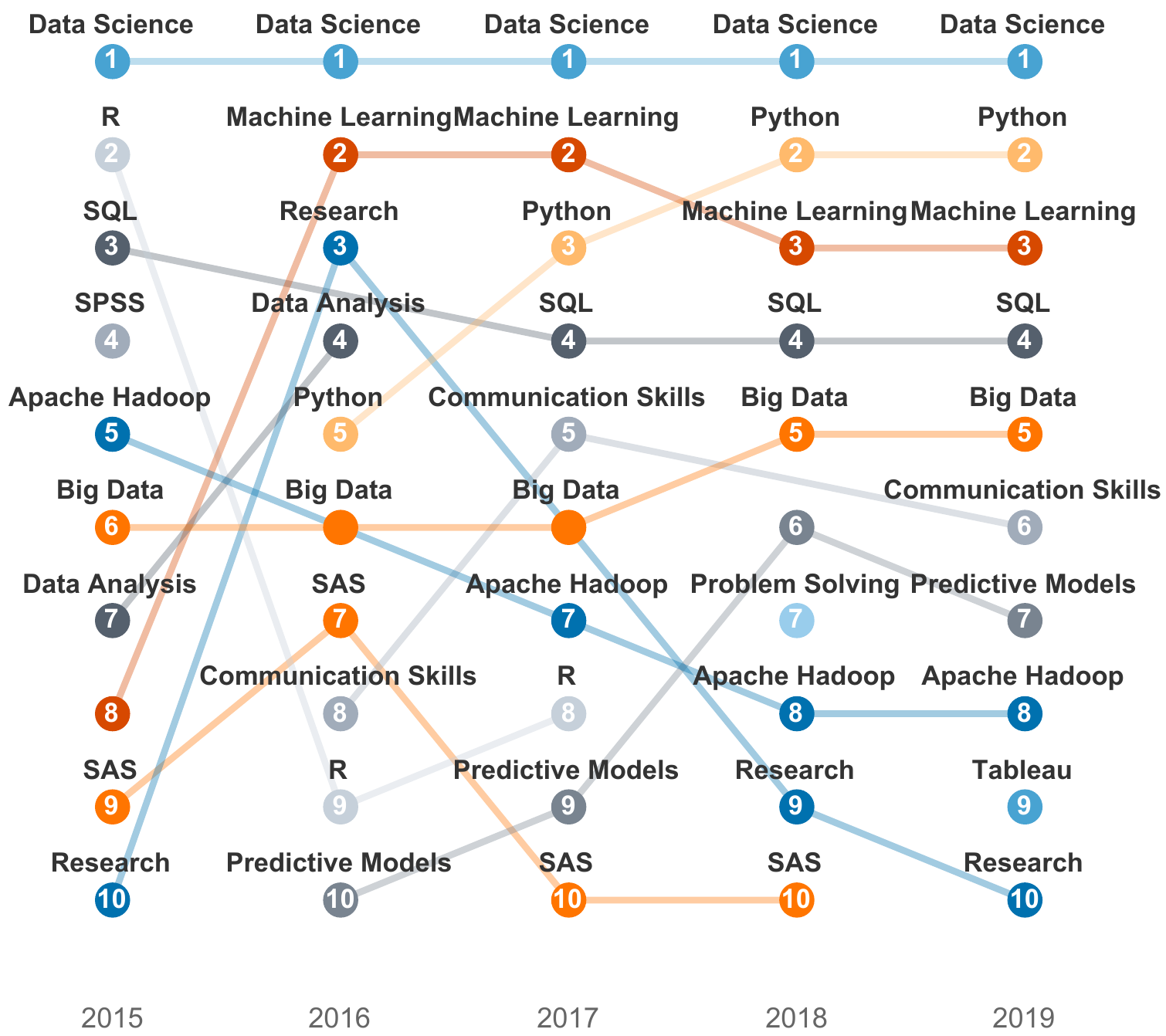}
		\label{subfig:ds-posting-freq}
	}
	\subfloat[]{
		\includegraphics[height = \myheightA\textheight]{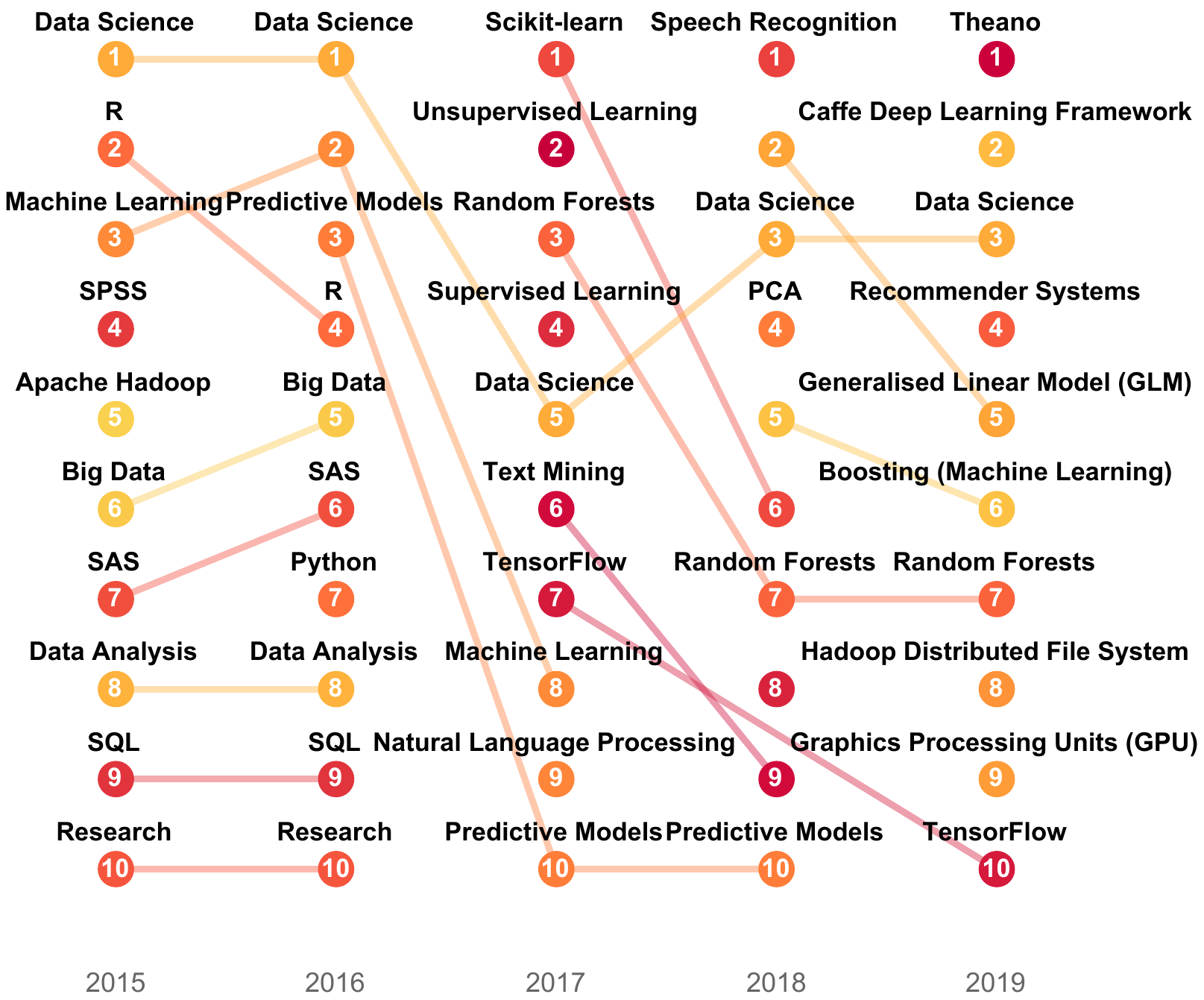}
		\label{subfig:ds-RCA}
	}
	\caption{
		\textbf{Comparison of two methods to analyze underlying skill demands of occupations in shortage}:
		\textbf{(a)} posting frequency of skills in an occupation; \textbf{(b)} Revealed Comparative Advantage of skills in an occupation to normalize highly-common skills and uncover skills most relevant to an occupation.
	}
	\label{fig:ds-skills}
\end{figure*}

\textbf{Most occupations are \textit{Not in Shortage}.}
In advanced labor markets, prolonged skill shortages are rare~\cite{brunello2019skill} \revA{and} 
most occupations are `Not in Shortage'.
\revA{This is visible in our ground truth data where}
there are over three times as many occupations classified as \textit{Not in Shortage} than \textit{In Shortage} \revA{(see \cref{subfig:total-shortage-dist})}. 
However, this has important modeling implications and requires hyper-parameter tuning to sufficiently adjust for these imbalances, as discussed in~\cref{subsec:predictive-model}.

\textbf{Some \revA{occupational} classes are \textit{In Shortage} more often than others.}
\revA{The shortage status of occupations is updated yearly in our dataset, and occupations can be \textit{In Shortage} for a period of time between 1 and 7 years (the extent of our dataset).}\mar{This paragraph was missing a description of what is represented in the figs, before interpretation. Added now.}
\revA{In \cref{subfig:yearly-shortage-dist}, we count the number of occupations \textit{In Shortage} based on the period of time they stay in shortage, and we color them by their occupational class.}
\revA{We observe} that \revA{the occupations belonging to the} `Technicians and Trades' \revA{class (shown in red)} are \textit{In Shortage} \revA{for longer periods of time than any} other occupational classes, including `Professionals'.
\revA{Furthermore, the `Technicians and Trades' class makes up the majority of occupations \textit{In Shortage} for four years or more.}
Generally, a small number of `Technicians and Trades' occupations classified \textit{In Shortage} tend to persist over several years, further illustrated in \cref{subfig:top-shortage-occs}. 
These finding, coupled with \revA{the fact that `Technicians and Trades' is the largest represented class in the ground truth (see}
\cref{subfig:major-group-dist}) \revA{indicates} that the ground-truth exhibits biases toward the `Technicians and Trades' workers occupational class \revA{-- probably due to the necessities of the Australian labor market}.

\textbf{Changes in labor shortage status.}
Changes to skill shortages of occupations are a key factor that determine adjustments to education, skilled immigration, and labor market policies. The ability to predict such yearly classification changes is therefore critical to models attempting to predict skill shortages.\mar{Again, what we measure before what we see.}
In \cref{subfig:total-shortage-dist}, \revA{we count the number of occupations \textit{In Shortage} and \textit{Not In Shortage} per each calendar year between 2012 and 2018, alongside with the number of occupations that flip their shortage status (from \textit{In Shortage} to \textit{Not In Shortage}, or the other way around, shown by the orange hexagrams).}
We see that changes to occupational skills shortage status are relatively rare \revA{(about 20 or less occupations change their status every year)}. 
This suggests that the ground-truth contains auto-regressive properties \revA{-- i.e., the status this year is most likely the same as last year --} which is an important modeling consideration, particularly for predicting shortage changes. 

\begin{figure*}
	\centering
	\newcommand\myheightA{0.20} 

	\subfloat[]{
		\includegraphics[height = \myheightA\textheight]{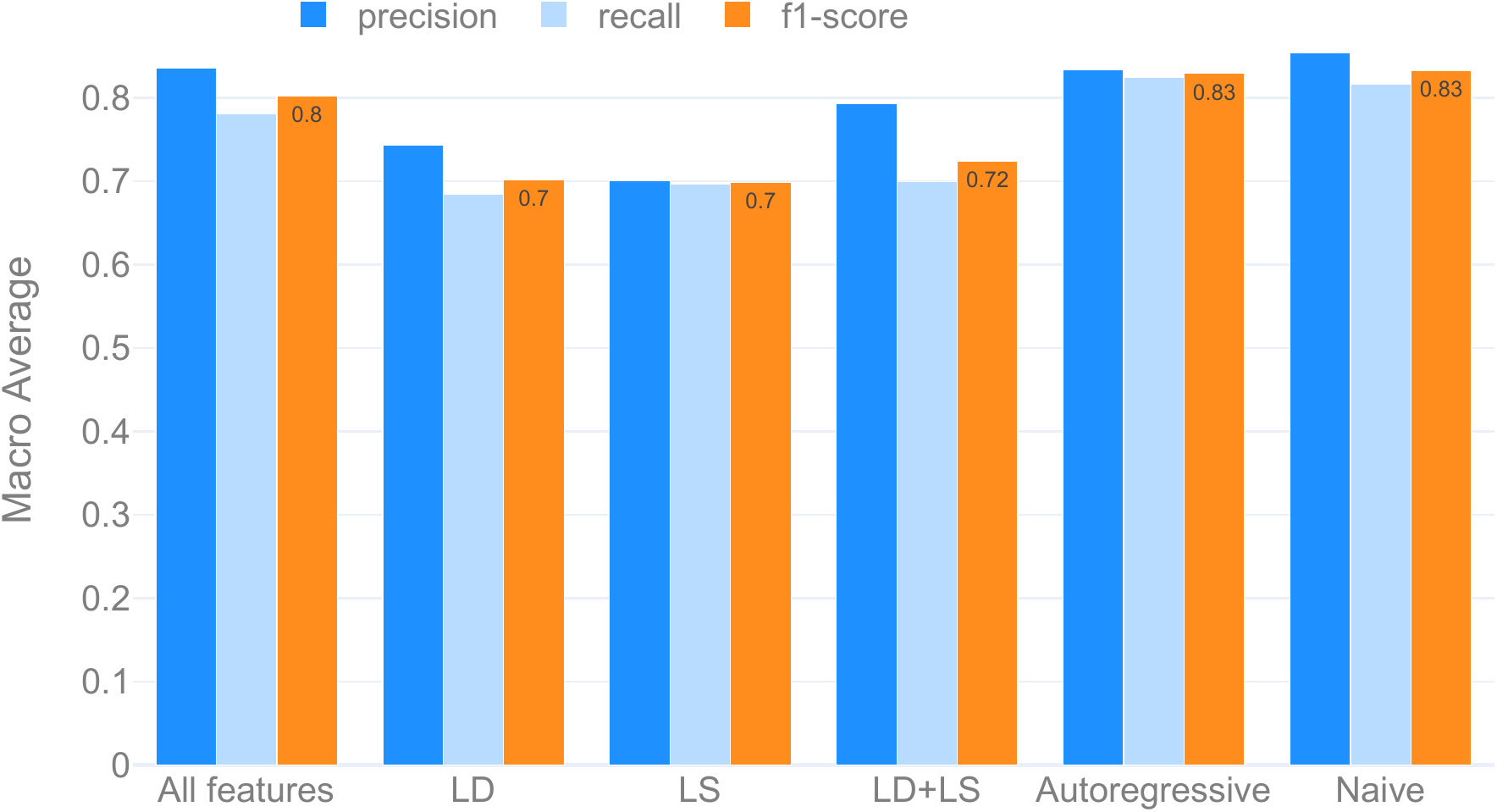}
		\label{subfig:all-occs-preds}
	}
	\subfloat[]{
		\includegraphics[height = \myheightA\textheight]{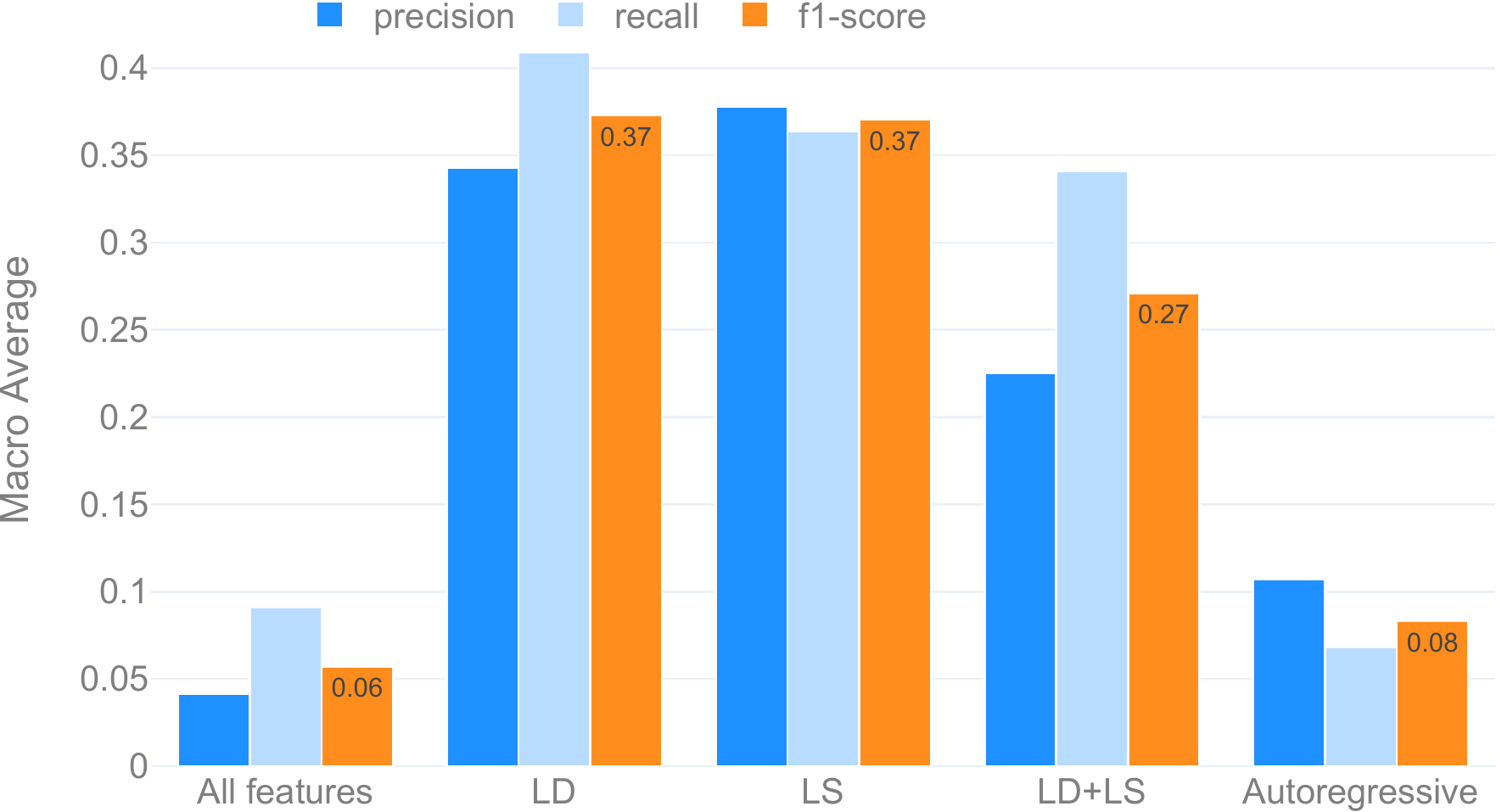}
		\label{subfig:occ-changes-preds}
	}
	\caption{
		\textbf{Skills Shortage prediction results}:
		\textbf{(a)} While the prediction results are highly auto-regressive, Labor Demand and Labor Supply features alone (and combined) perform almost as well for predicting occupational shortages; \textbf{(b)} Labor Demand and Labor Supply features perform better than other features at predicting shortage status changes of occupations.
	}
	\label{fig:shortage-preds}
\end{figure*}

\subsection{Skill importance levels for Data Scientists.}
\label{subsec:data-scientists-skill-importance}
\mar{Some content here was duplicated from \mbox{\cref{subsec:skill-importance-methods}}. Shortened!}%
\revA{Here, we compare the two approaches to assess skill importance for occupations that we introduced in \cref{subsec:skill-importance-methods}, and we showcase them for the occupation `Data Scientist'.}
\revA{The first approach}
is to perform temporal skill counts grouped by occupation (or another grouping source). 
\cref{subfig:ds-posting-freq} highlights \revA{the top 10 skills for `Data Scientist' obtained using this approach, for each between 2015 and 2019.
Visibly, skills like} `Communication Skills', `Research', and `Problem Solving' regularly rank in the top 10, however, these are among the most common skills in the BGT dataset -- for example, `Communication Skills' is present in over one-quarter of all job ads.
\revA{This is because skill counts do} not normalize for highly common skills that are present in all or most occupations, making it
questionable whether this can be used as a proxy for skill importance within an occupation.

\revA{The second approach detailed in \cref{eq:SSI} is the RCA approach.}
\cref{subfig:ds-RCA} \revA{shows the} top 10 yearly skills \revA{obtained using RCA.
Visibly, the obtained top skills are} considerably more specific to the `Data Scientist' occupation. 
Machine Learning and Deep Learning tools and techniques dominate the ranked list, while some core Data Science skills seen in \cref{subfig:ds-posting-freq} remain. 
\revA{This method also captures the rise of emerging skills (such as \emph{Generalized Linear Models}, \emph{Boosting} or \emph{Random Forests}), which}
are critical for occupations in shortage.

\subsection{Predict Skill Shortages}
\label{subsec:prediction-results}
\mar{disabled entirely the discussion about the number of lagged features, as already mentioned \cref{subsec:predictive-model}.}

\revA{Here, we detail the results of two predictive exercises.
First, we predict the shortage status of occupations and we perform an ablation study to identify the most important sets of features.
Second, we show the results of the more difficult task of predicting shortage status changes (when an occupation flips its shortage status between \textit{In Shortage} and \textit{Not In Shortage}).}

\textbf{\revA{Predict occupation shortages.}}
\revA{We predict occupation shortages following the setup described in \cref{subsec:predictive-model}.
We equally study which class of features is most predictive by performing an ablation study -- i.e. we repeat the predictive experiment multiple times using all the features, or only subsets of features.}
We train and evaluate the following feature input configurations:
\textbf{All-In}: all features included;
\textbf{LD}: labor demand features included only;
\textbf{LS}: labor supply features included only;
\textbf{LD + LS}: labor demand and labor supply features included;
\textbf{Auto-regressive Predictor}: lagged target features included only;
\textbf{Naive Predictor}: copy target variable from the previous time period.
%
\revA{\cref{subfig:all-occs-preds} shows the prediction performance -- macro- precision, recall and F1 (higher is better) -- of the different setups.
Due to the strong auto-regressive properties of the problem,}
the \textit{Naive Predictor} and the \textit{Auto-regressive Predictor} achieve the highest performance ($F1 = 83\%$), \revA{however they always predict the last shortage status for each occupation.
These predictors are useless for occupations that flip their status, which are of strong interest in real-world applications. Visibly, the models that exclude the auto-regressive features (i.e. the LD and/or LS models), maintain solid performance levels (up to F1=72\%). The significance of this finding is discussed in \cref{sec:discussion}.} 
%
%

\textbf{Predict shortage \revA{status} changes.}
\revA{We evaluate the same classifiers trained as described above on a slightly different problem: how well can they predict the \emph{changes in shortage status}?}
To achieve this, we filtered occupations in the testing dataset to include only those with a different skills shortage classification to the previous year.
For example, as `Architects' were classified as \textit{In Shortage} in 2017 but were \textit{Not in Shortage} in 2016, they were therefore included in the performance evaluation.
\revA{\cref{subfig:occ-changes-preds} shows the resulting prediction performance: precision, recall and F1.
Visibly, the performances decreased substantially, and the hardest hit are the models leveraging the auto-regressive property (with \textit{Naive} obtaining zero everywhere).
The reason is that shortage status changes} are fairly rare events, (see \cref{subfig:total-shortage-dist}) \revA{which auto-regressive classifiers completely miss.}
The highest performing models use LD or LS features. This is particularly relevant to real-world scenarios, where researchers closely follow occupations that change their status, as this has policy and immigration implications.

\subsection{Feature Importance for Predicting Skill Shortages}
\label{subsec:feature-importance}
As seen in \cref{subfig:all-occs-preds}, the model with the auto-regressive features has the highest performance, so previous shortage classifications are the most important features for predicting skill shortages in this dataset. However, longitudinal datasets of skill shortages, like the data used for this analysis, are rare. Therefore, auto-regressive target features are often unavailable for analyzing skill shortages in other labor markets. Labor demand and labor supply features, however, are standard and available across most labor markets. 
\revA{Here}, we conduct feature importance analysis on the `LD + LS' model seen in \cref{subfig:all-occs-preds} in order to draw insights into which of these features are most predictive of skill shortages. We use the `Gain' metric, which shows the relative contribution of each feature to the model by calculating the features' contribution for each tree in the \revA{XGBoost} model. A higher gain score indicates that a feature is more important for generating a prediction. 

\cref{fig:feature-importance} shows that variations of \revA{the labor supply feature} `Hours Worked' are the most predictive for skill shortages, \revA{as they}
account for 6 of the top 20 most important features \revA{-- see positions 1, 2, 4, 10, 12, 15 in \cref{fig:feature-importance}}. 
The next most important features belong to the labor demand class. Namely, years of `Education' and `Experience' demanded by employers and median `Salary' levels in job ads. A brief interpretation of these feature importance levels follows in \cref{sec:discussion}.

\begin{figure}[tbp]
    \centering
    \includegraphics[width=0.48\textwidth]{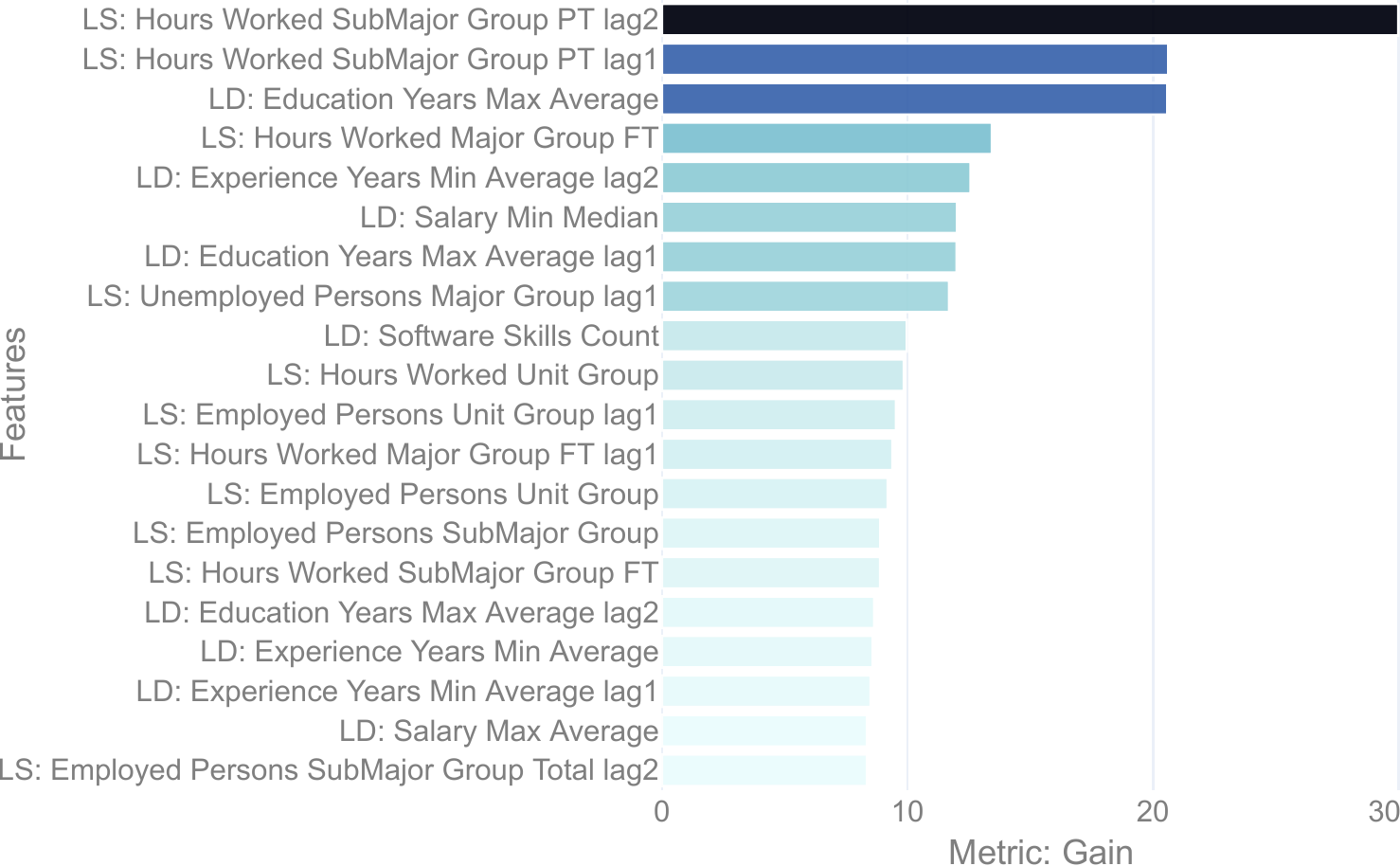}
    \caption{
    	Feature importance of Labor Demand and Labor Supply feature model.
    }
    \label{fig:feature-importance}
\end{figure}

\section{Discussion}
\label{sec:discussion}
\textbf{\revA{Trade-off between performance and data availability.}}
The highest performing models in \cref{subfig:all-occs-preds} exhibit strong auto-regressive properties. 
This is to be expected given that changes in the skill shortage status of occupations tend to be rare, as seen in \cref{subfig:total-shortage-dist,subfig:top-shortage-occs}. 
However, removing the auto-regressive target features and leaving the labor demand and labor supply features maintains a relatively strong result (F1=72\%). 
This is significant because \revA{labor demand and labor supply} data sources \revA{are available}
across multiple labor markets, \revA{whereas}
longitudinal skill shortages data at the occupational level \revA{are rare} in most labor markets. 
This suggests that while labor demand and labor supply data contain rich information for detecting skill shortages, there is a trade-off between prediction performance and data availability when deploying in new labor markets.

\textbf{\revA{Auto-regressive features cannot predict shortage status changes.}}
\revA{Given the strong auto-regressive nature of skill shortages (i.e. the best indicator of an occupation being in shortage this year is if it was in shortage last year), classifiers have the tendency of over-leveraging the information from the past.
While this may help performance indicators, it considerably reduces the value of the prediction in a real-life setups.
Shortage status changes (when an occupation moves between \textit{Not In Shortage} and \textit{In Shortage}) have policy and immigration implications, as governments decide skilled immigration rules based on the needs of the labor market.
In other words, it is more important to be able to predict when an occupation shortage \emph{status changes} than simply predicting its next status.
Visibly in \cref{subfig:occ-changes-preds}, the performances of the classfiers leveraging auto-regressive features are significantly reduced when predicting shortage status changes.}
%
Nonetheless, we found that labor demand and labor supply data were most predictive of shortage changes, respectively. This is significant because it further highlights the value of near-real time data sources (job ads data) and freely available data sources (employment statistics). 
%
Both labor demand and labor supply data sources could be leveraged to replicate our modeling approach in other markets to assist policy-makers to better preempt skill shortage changes of occupations. This could help with critical tasks such as forward planning for education and training policies, skilled immigration, and workforce transitions.

\textbf{Understanding what matters for predicting skill shortages.} The most important features from the `LD + LS' model (seen in \cref{fig:feature-importance}) are consistent with the literature on skill shortages~\cite{brunello2019skill, healy2015, Richardson2007-vl, cedefop2018insights, OECD2019Skills, Dawson2019}. 
Specifically, `Hours Worked' is considered an important indicator for occupations in shortage~\cite{healy2015, Richardson2007-vl} \revA{due to the following rationale:} 
when a shortage exists for an occupation, the demands placed upon workers classified in that occupation are naturally high, which manifests in higher work intensity and longer work hours. 
This is reflected in \cref{fig:feature-importance} where the `Hours Worked' variables are represented in 6 of the top 20 most important features. 

With regards to labor demand, years of `Education', years of `Experience', and median `Salary' are all highly important features for predicting occupational skill shortages. 
This is consistent with prior work~\cite{Dawson2019}, \revA{which shows} that when an occupation is in shortage, employers adjust job requirements to try and fulfill their demands. 
With regards to these features, this typically involves lowering the requirements of education and experience and increasing salary levels to attract more candidates.

\section{Conclusion and Future Work}
\label{sec:conclusion}
In this research, we (1) compared two methods to analyze the skill demands of occupations in shortage; (2) we constructed a Machine Learning framework to predict temporal skill shortages of occupations; and (3) we analyzed feature importance data to understand which labor supply and labor demand features are most predictive of occupational skill shortages. The methods and findings from this work can assist policy-makers to better measure and predict skill shortages of occupations. Similarly, educators could apply this work to better identify market demands and adjust their curricula accordingly.

The biggest limitation with skills shortage research is the lack of representative data at the occupational level. The `Historical List of Skills Shortages in Australia', used in this research, is among the world leaders in this regard, despite its shortcomings discussed in \cref{sec:data-methods}. 
Therefore, systematically measuring occupational skill shortages is arguably the most important work that can be done to advance the knowledge of skill shortages. 
Other future work could apply the framework we have developed predict skill shortages in other labor markets. Additionally, different features could be constructed as descriptive variables, and more auto-regressive lag periods could be considered. Lastly, another research avenue could assess how these results could be improved by applying other predictive tools, such as Deep Learning approaches.


\Urlmuskip=0mu plus 1mu\relax

{\small
\bibliographystyle{plain}
\bibliography{bibliography}
}

%
%
\clearpage
\appendix
\etocdepthtag.toc{mtappendix}
\etocsettagdepth{mtchapter}{none}
\etocsettagdepth{mtappendix}{subsection}
\etoctocstyle{1}{Contents (Appendix)}
\tableofcontents

\section{Appendix}

This document is accompanying the submission \textit{\titlename}.
The information in this document complements the submission, and it is presented here for completeness reasons.
It is not required for understanding the main paper, nor for reproducing the results.

\subsection{Cyclical and Structural Factors Affecting Skill Shortages}
\label{subsec:cyclic-structural-factors}

\textbf{Macroeconomic cycles} can affect skill shortages. 
During periods of economic expansion, skill shortages tend to increase as firms seek to hire skilled labor to meet new and growing market demands~\cite{brunello2019skill}.
The `Manpower Talent Shortage Survey'~\cite{ManpowerGroup2018-lb} is the largest skill shortage survey in the world. 
The global survey found that skill shortages have increased from 30\% in 2009 to 45\% in 2018, equating to a 12 year high.
Similarly, the annual Cedefop skills mismatch survey in Europe~\cite{cedefop2018insights} found that labor market shifts in the aftermath of the economic crisis have resulted in the stated inability of employers to fill their vacancies with suitably skilled workers.

\textbf{Structural changes} to labor markets also influence skill shortages. 
These most notably take the form of demographic changes, technological advances, and globalization. 
Demographic changes affect the demand for goods and services. 
For instance, as the average age of a population increases, so does their demand for healthcare services. 
This subsequently increases the aggregate labor demand for workers with healthcare related skills~\cite{OECDHealth}.
As the average age is increasing for almost all advanced economies~\cite{WHO}, these structural demographic changes are likely to affect skill shortages for specific occupational classes, such as healthcare services.

\textbf{Technological advances} introduce structural changes that can exacerbate skill shortages. 
As firms adopt new technologies, they seek skilled labor to implement and make productive use of these new technologies. 
This can create dynamics of `skill biased technological change'~\cite{katz1999, Autor2011}, whereby the acceleration of demand for technical skills outweighs the available supply of workers who possess such skills. 
There is evidence of these dynamics currently occurring as a result of the growing demands for Data Science and Machine Learning skills~\cite{Dawson2019}.
While the capacity to collect, store, and process information may have sharply risen, it is argued that these advances have far outstripped present capacities to analyze and make productive use of such information~\cite{Hey2009-hj}.
Claims of Data Science and Advanced Analytics (DSA) skill shortages are being made in labor markets around the world~\cite{Blake2019-ha,LinkedIn_Economic_Graph_Team2018-gu,Manyika2011-vh}.
Two studies conducted using job ads data assessed DSA labor demands and the extent of skill shortages. 
The first was an industry research collaboration between Burning Glass Technologies (BGT), IBM, and the Business-Higher Education Forum in the US~\cite{Markow2017-hu}. 
The research found that in 2017 DSA jobs earned a wage premium of more than US\$8,700 and DSA job postings were projected to grow 15\% by 2020, which is significantly higher than average. 
In another study commissioned by the The Royal Society UK~\cite{Blake2019-ha}, job ads data were analysed for DSA jobs in the UK. 
The results again also showed high and growing levels of demand for DSA skills (measured through posting frequency) and wage premiums for DSA related occupations.

\textbf{Globalization} can act as a shock to labor markets that induce or deepen skill shortages. 
The offshoring of labor tasks can increase the polarization of labor markets by reducing the domestic demand for middle-skilled jobs~\cite{brunello2019skill}.
This causes a process of labor reallocation, as workers attempt to transition between jobs. 
If the reallocation of labor is inefficient, skill shortages can increase because the supply of skilled workers is insufficient to meet the evolving labor demands of growing sectors.

\textbf{The current work} proposes a robust data-driven method that assesses skill shortages and that uses machine learning to account for the factors that affect skill shortages.

\subsection{Oversampling}
Oversampling is a technique that involves randomly duplicating observations from the minority class (\textit{In Shortage} in the case of this research) and adding them to the training dataset.
The main benefit of oversampling is that it creates a balanced distribution of target variables without `data leakage' that occurs from `under-sampling' (that is, randomly removing observations from the majority class). 
Creating a balanced distribution of predictive classes is particularly important for a range of classification algorithms~\cite{Branco2015}.
However, a shortcoming of oversampling is that it can increase the likelihood of overfitting, as exact copies of the minority class are constructed~\cite{Fernandez2018}.
The oversampling ratio is defined as:
\begin{equation*}
	Oversampling\ Ratio = \frac{\mathop{\sum}(Majority\ Class)}{\mathop{\sum}(Minority\ Class)}
\end{equation*}

The output of this ratio was specified as a hyper-parameter value in each model type that we constructed. 

\subsection{Performance metrics}
Precision measures how many of the predictions were correct. 
Recall measures the completeness of the prediction -- how many of the true answers were correctly uncovered. 
The F1 is the harmonic mean of precision and recall -- a classifier needs to achieve both a high precision and a high recall in order to obtain a high F1.
Formally, these are defined as:
\begin{equation*}
	Precision = \frac{TP}{TP + FP};
\end{equation*}

\begin{equation*}
	Recall = \frac{TP}{TP + FN};
\end{equation*}

\begin{equation*}
	F1 = \frac{2 \cdot Precision \cdot Recall}{ Precision + Recall}
\end{equation*}

where $TP$ are the number of true positives -- number of correctly identified items of the class of interest; 
$FP$ are false positives (items incorrectly predicted as pertaining to the class of interest);
and $FN$ are false negatives (items incorrectly predicted as not being of interest).
Note that one can compute the precision, recall and F1 for each class of interest (here, both the \emph{In Shortage} and the \emph{Not in Shortage}), and the scores for each class could be wildly different as one class might be more predictive than the other.

\subsection{Using a standardized occupation taxonomy -- ANZSCO.}
All data sources mentioned above correspond to their respective occupational classes according to the Australian and New Zealand Standard Classification of Occupations (ANZSCO).~\cite{ANZSCO2013}
ANZSCO provides a basis for the standardized collection, analysis and dissemination of occupational data for Australia and New Zealand. 
The structure of ANZSCO has five hierarchical levels - major group, sub-major group, minor group, unit group and occupation. 
The categories at the most detailed level of the classification are termed 'occupations'. 
Depending on data availability, labor statistics were included in the models from the occupation level through to the major group level.

There are some significant shortcomings to analyzing occupations within ANZSCO classifications. 
Official occupational classifications, like ANZSCO, are often static taxonomies and are rarely updated. 
They therefore fail to capture and adapt to emerging skills, which can misrepresent the true labor dynamics of particular jobs. 
For example, a `Data Scientist' is a relatively new occupation that has not yet received its own ANZSCO classification. 
Instead, it is classified as an `ICT Business \& Systems Analyst' by ANZSCO, grouped with other job titles like `Data Analysts', `Data Engineers', and `IT Business Analysts'. 
However, as ANZSCO is the official and prevailing occupational classification system, all data used for this research are in accordance with the ANZSCO standards.

\subsection{Summary of constructed features}
\begin{table*}
    \centering
    \caption{
    	Summary of constructed features and their explanation.
    }
    {\small
    \begin{tabular}{crp{8.3cm}}
        \toprule
        & Name & Meaning and explanation \\ \midrule
        \multirow{12}{*}{\rotatebox[origin=c]{90}{Labour Demand}} & Posting Frequency: & number of job advertisement vacancies \\
        & Max Median Salary: & maximum median salary advertised \\
        & Min Median Salary: & minimum median salary advertised \\
        & Max Average Salary: & maximum average salary advertised \\
        & Min Average Salary: & minimum average salary advertised \\
        & Max Average Experience: & maximum average years of experience required \\
        & Min Average Experience: & minimum average years of experience required \\
        & Max Average Education: & maximum average years of formal education required \\
        & Min Average Education: & minimum average years of formal education required \\
        & Specialised Count: & total count of required skills considered specialised to a specific vocation \\
        & Baseline Count: & total count of skills that are considered applicable across vocations  \\
        & Software Count: & total count of skills that are software-related \\
       \midrule
       \multirow{22}{*}{\rotatebox[origin=c]{90}{Labour Supply}} & Unit Total Employed: & total number employed at ANZSCO Unit level (000's)\\
        & Unit Total Hours Worked: & total hours worked at ANZSCO Unit level (000's)\\
        & Sub FT Employed: & total employed full-time at ANZSCO Sub-Major level (000's)\\
        & Sub PT Employed: & total employed part-time at ANZSCO Sub-Major level (000's)\\
        & Sub Total Employed: & total employed at ANZSCO Sub-Major level (000's)\\
        & Sub FT Hours Worked: & total full-time hours worked at ANZSCO Sub-Major level (000's)\\
        & Sub PT Hours Worked: & total part-time hours worked at ANZSCO Sub-Major level (000's)\\
        & Sub Total Hours Worked: & total hours worked at ANZSCO Sub-Major level (000's)\\
        & Major FT Employed: & total employed full-time at ANZSCO Major level (000's)\\
        & Major PT Employed: & total employed part-time at ANZSCO Major level (000's)\\
        & Major Total Employed: & total employed at ANZSCO Major level (000's)\\
        & Major FT Hours Worked: & total full-time hours worked at ANZSCO Major level (000's)\\
        & Major PT Hours Worked: & total part-time hours worked at ANZSCO Major level (000's)\\
        & Major Total Hours Worked: & total hours worked at ANZSCO Major level (000's)\\
        & Major Unemployed FT Seekers: & total unemployed seekers full-time at ANZSCO Major level (000's)\\
        & Major Unemployed PT Seekers: & total unemployed seekers part-time at ANZSCO Major level (000's)\\
        & Major Unemployed Total Seekers: & total unemployed seekers at ANZSCO Major level (000's)\\
        & Major Total Weeks Searching: & total number of weeks unemployed persons job searching at ANZSCO Major level (000's)\\
        & Major Underemployed Total: & total number of persons underemployed at ANZSCO Major level (000's)\\
        & Major Underemployed Ratio: & ratio of underemployed persons at ANZSCO Major level \\
        \bottomrule
    \end{tabular}
    }
    \label{tab:feature-summary}
\end{table*}

\subsection{Feature correlation analysis}
\begin{figure*}
    \centering
    \includegraphics[width=16cm]{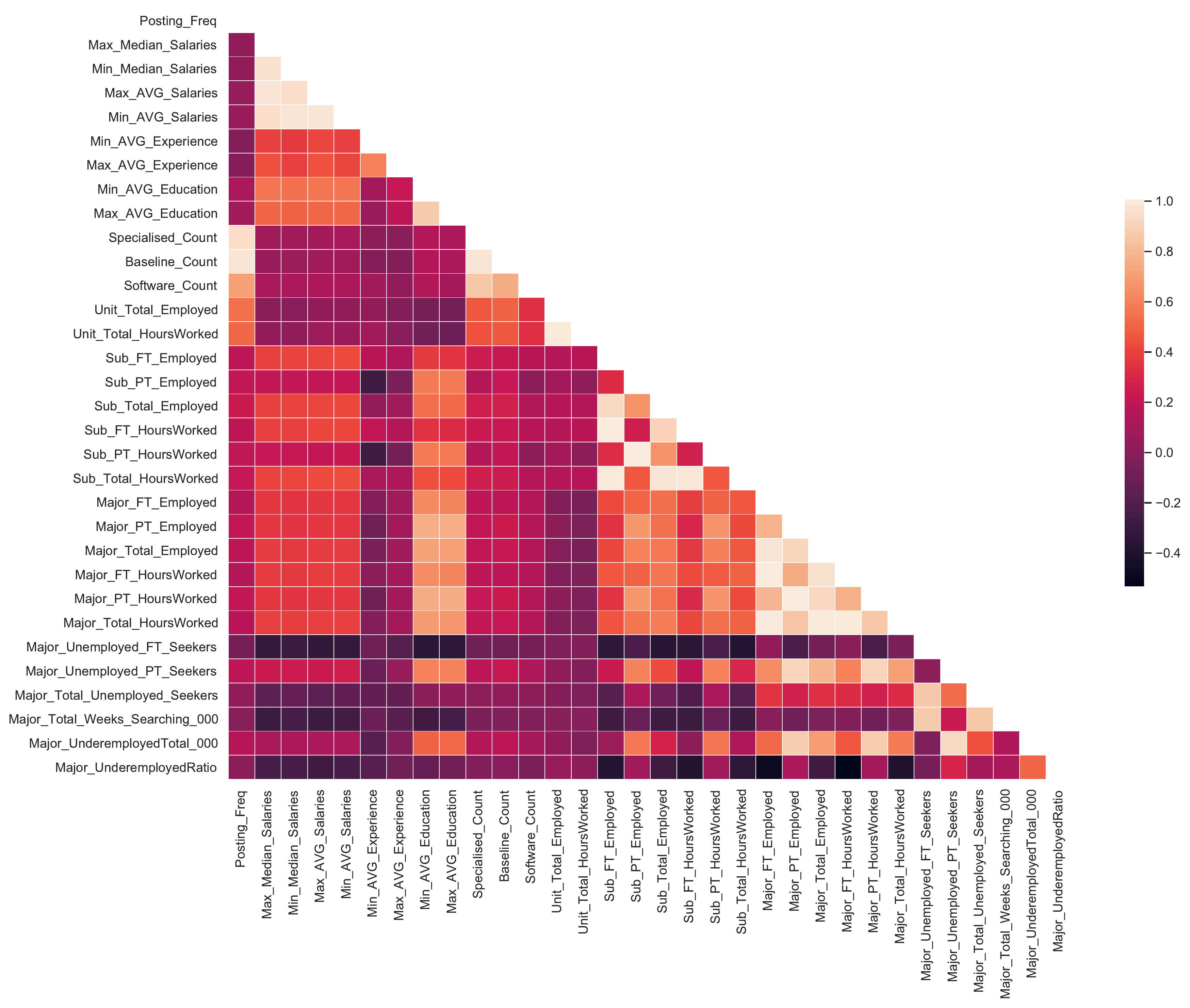}
    \caption{Correlation analysis between modeled features.}
    \label{fig:feature_corr}
\end{figure*} 

\end{document}